\documentclass{article}
\usepackage{graphicx}
\usepackage{color}
\usepackage{newcent}
\usepackage{microtype}
\usepackage[a4paper]{geometry}

\title{\huge \sf Conversational Sensing}
\author{Alun Preece\thanks{Revised preprint of ``Conversational Sensing'' in {\it Next Generation Analyst II, SPIE DSS 2014}. Send correspondence to PreeceAD@cardiff.ac.uk}\\
Chris Gwilliams\\
Christos Parizas\\
Diego Pizzocaro\\
School of Computer Science and Informatics\\
Cardiff University, Cardiff, UK \\ \\
Jonathan Z. Bakdash\\
Human Research and Engineering Directorate\\
US Army Research Laboratory\\
Aberdeen Proving Ground, USA \\ \\
Dave Braines\\
Emerging Technology Services\\
IBM United Kingdom Ltd\\
Hursley Park, Winchester, UK}

\date{}

\begin{document} 
\maketitle 

\begin{abstract}
Recent developments in sensing technologies, mobile devices and context-aware user interfaces have made it possible to represent {\color{black} information fusion and situational awareness} as a {\it conversational} process among actors --- human and machine agents --- at or near the tactical edges of a network. Motivated by use cases in the domain of {\color{black} security, policing and emergency response}, this paper presents an approach to information collection, fusion and sense-making based on the use of natural language (NL) and controlled natural language (CNL) to support richer forms of human-machine interaction. The approach uses a conversational protocol to facilitate a flow of collaborative messages from NL to CNL and back again in support of interactions such as: turning eyewitness reports from human observers into actionable information (from both {\color{black} trained and untrained sources}); fusing information from humans and physical sensors (with associated quality metadata); and assisting human analysts to make the best use of available sensing assets in an area of interest (governed by management and security policies). CNL is used as a common formal knowledge representation for both machine and human agents to support reasoning, semantic information fusion and generation of rationale for inferences, in ways that remain transparent to human users. Examples are provided of various alternative styles for user feedback, including NL, CNL and graphical feedback. A pilot experiment with human subjects shows that a prototype conversational agent is able to gather usable CNL information from untrained human subjects.
\end{abstract}

\section{Conversational D2D}
\label{sec:intro}

\begin{figure}
\centering
\includegraphics[width=0.75\textwidth]{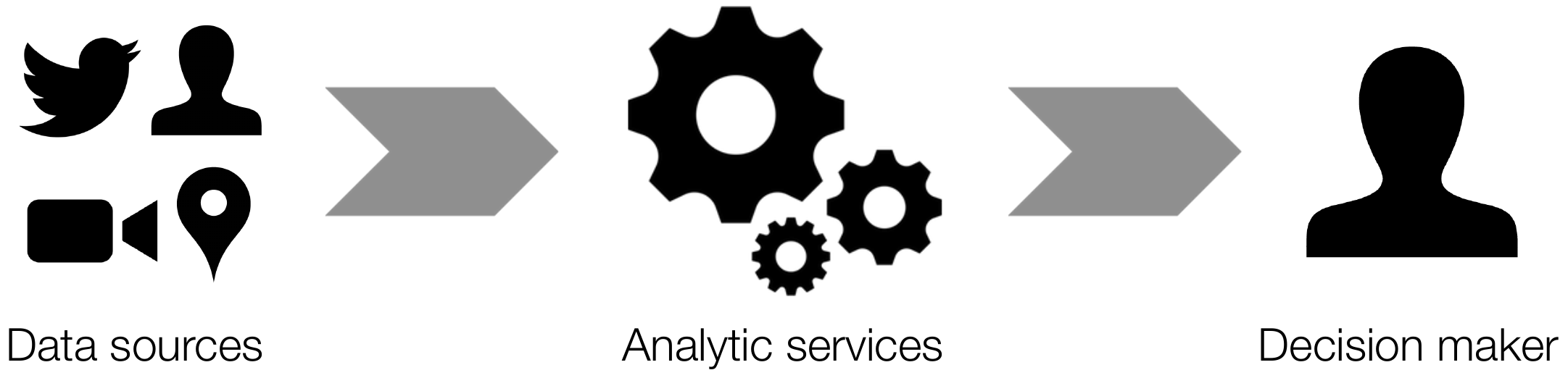}
\caption{An abstract data-to-decision pipeline}
\label{fig:d2d1}
\end{figure}

We live in an age where unprecedented amounts of data are available to inform human decision-making. In the UK, {\it big data} has been identified as the first of ``eight great technologies'' for economic growth\footnote{http://www.policyexchange.org.uk/images/publications/eight\%20great\%20technologies.pdf}. In the US, the Department of Defense listed its first science and technology priority for 2013--17\footnote{http://www.acq.osd.mil/chieftechnologist/publications/docs/OSD\%2002073-11.pdf} as {\it data to decisions} (D2D): ``Science and applications to reduce the cycle time and manpower requirements for analysis and use of large data sets''~\cite{Broome2012}. The wording here emphasises that the data landscape is changing rapidly and, to be effective, ``big data'' techniques --- including data collection, analytics and visualisation --- need to be highly agile. The typical model for a D2D pipeline is shown in Figure~\ref{fig:d2d1}, where data are collected from one or more data sources of various kinds, processed by a variety of analysis services and the results delivered to the decision-maker in some actionable form. 

Available data sources --- often characterised by the ``three Vs'', volume, velocity and variety~\cite{Laney:2001} --- span an enormous range of types, including physical sensors, geospatial and other information systems, social media of many kinds and human sources. Often it is necessary to combine data from multiple heterogeneous sources, through some process of information fusion~\cite{jdl:Llinas:2004}. Analytic services are equally diverse, including signal processing, statistical, machine learning and inferential systems. Again, often multiple analytic processes are used in combination, for example signal processing to identify lower-level features, followed by inference to perform higher-level classification. The optimal form for information retrieval and delivery to a human decision-maker depends on both human capabilities and system capabilities. Contrary to intuition, providing more information does not necessarily improve human decision-making~\cite{Bakdash:2013,Goldstein:2002,Hall:2007}. Thus, a gist-level representation of information, with the ability to drill-down to see rationale and supporting evidence, is key to supporting effective human decision-making. The physical hardware for accessing the system is another consideration for the form of information: for example, delivery to a mobile user via a smartphone app or wearable device requires different human-computer interaction approaches than delivery to a large, conventional display screen. 

In addition to work on techniques for data collection, processing and dissemination, there has also been significant investment in tools and methods to make it easier and quicker for developers to construct D2D pipelines, including research in middleware, platforms and automated workflows~\cite{Dumbill:2012}. However, the majority of work in this space has taken a data-driven view. A problem that has received less attention is how to rapidly construct pipelines by working backwards from an intended decision (or hypothesis or query) and identifying useful analysis services and underlying data sources that can assist the decision-maker~\cite{Geyik:2013,Preece:2013}. The collection and availability of information is necessary, but not sufficient for assisting the decision-maker. For optimal decision-making, the (human) search costs must be minimized; that is, the decision-maker must be able to access information in a timely manner~\cite{Fu:2006}. 

\begin{figure}[h]
\centering
\includegraphics[width=0.79\textwidth]{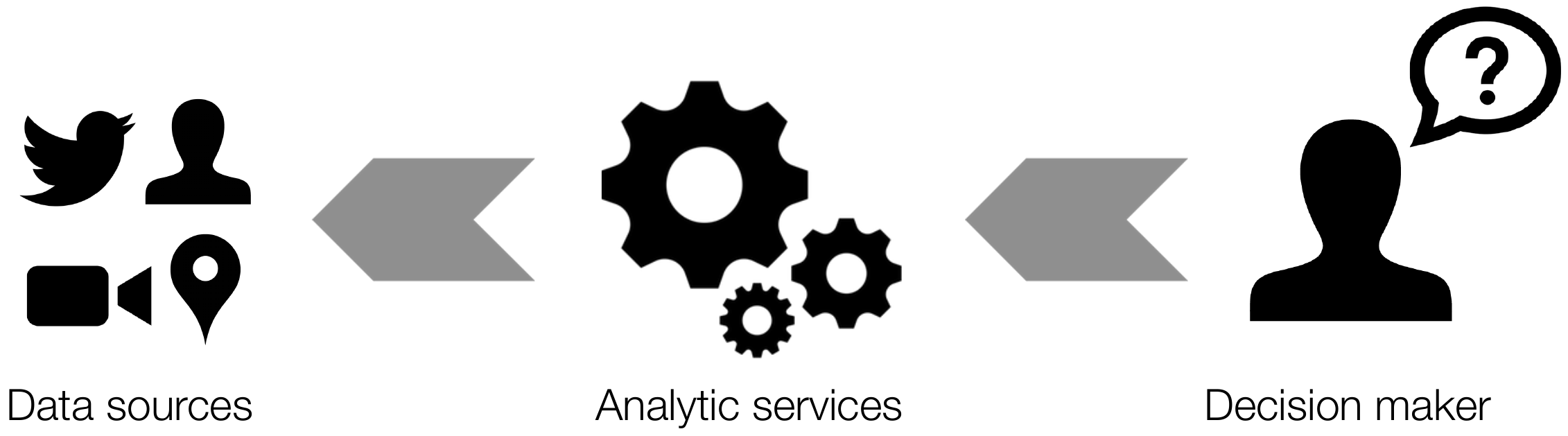}
\caption{Constructing a D2D pipeline dynamically by a ``backward chaining'' process}
\label{fig:d2d2}
\end{figure}

We see this in well-publicised incidents such as the damage to Japan's Fukushima nuclear plant in the wake of the 2011 earthquake. An urgent need arose to monitor radiation leaks --- the decision-maker's intent --- leading to the rapid deployment of networked Geiger counters, many of which were private devices shared via Internet of Things approaches\footnote{http://www.wired.com/opinion/2012/12/20-12-st\_thompson/}. This ``backward chain'' from decision intent to data sources is shown in Figure~\ref{fig:d2d2}. Note that the arrows here depict control flow: the user's intent frames the problem, leading to the selection of suitable services and compatible data sources. The result is a dynamically constructed pipeline as shown in Figure~\ref{fig:d2d1}. Building this pipeline on-the-fly through a highly automated process of service and source selection, and composition is a method to address the original priority to ``reduce the cycle time and manpower requirements'' in D2D. 

A number of trends seem to point towards an even more flexible and agile view of D2D systems. Firstly, the data sources are becoming increasingly ``smart'' and communicative. Autonomous vehicles and robotic systems, together with increasingly computationally capable Internet of Things devices operating in peer-to-peer networks open up greater potential for collective intelligence and self-organisation at what has traditionally been the edge of the network, where data sources are often located. At the same time, the rapidly increasing sophistication of mobile devices has freed decision-makers to operate in contexts much nearer the tactical edge. Mobile users have become adept at agile, on-the-fly decision-making, able to cope with dynamically changing sets of requirements while simultaneously carrying out actions in the field. Many activities previously seen as strategic or operational in decision-making terms, have been ``tacticalised'' by mobile technology. Pervasive information sources have given rise to a new generation of context-aware, assistive technologies typified by Apple's Siri\footnote{https://www.apple.com/ios/siri/} and Google Now\footnote{http://www.google.com/now/}. These technologies are changing the modes of interaction between users and their devices, with the device able to take an increasingly active role in the interaction, for example, by engaging in conversation or pushing notifications to the user in an anticipatory manner. 

\begin{figure}
\centering
\includegraphics[width=0.79\textwidth]{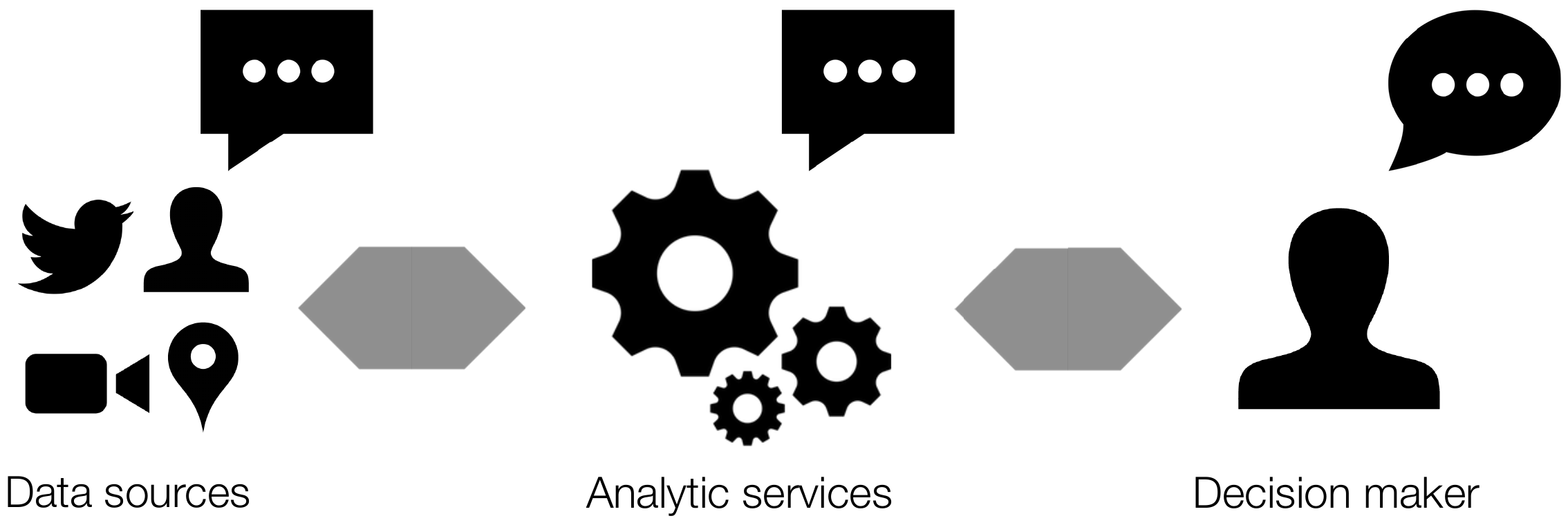}
\caption{Conversational D2D}
\label{fig:d2d3}
\end{figure}

In this context, the traditional D2D pipeline can be re-thought as a collection of interactions between agents with different specialisms: the data sources, analytic services and decision-makers can be viewed as engaging in peer-to-peer interactions with each other, with chains of interaction able to start anywhere in the network and flow in any direction, from data to decision, or from query to response. This ``conversational D2D'' model is shown in Figure~\ref{fig:d2d3}. The different styles of ``speech bubble'' here suggest that the machine participants will tend to communicate in structured message formats, while the human participants will use a more natural form of interaction --- this topic is developed in Section~\ref{sec:appro}.

To ground this rethinking of the D2D pipeline in some concrete examples, consider a number of use cases in the context of {\color{black} typical field situations involving reports from human observers, information fusion and inference, and the tasking of assets. These examples can be apply to policing (especially in a non-urban context), border security, or environmental protection (for example, countering poaching).}

\subsection{Spot Report}

A human in a particular location generates an eyewitness (``spot'') report using their mobile device. For example, they may report a suspicious vehicle. Here the human is acting initially as a data source: on the left of Figure~\ref{fig:d2d3}. If they are trained in doing this (e.g. {\color{black} they are a police officer, soldier, or wildlife warden}) the person will probably use a structured format; in other cases their report will be unstructured. In either case, there will be things from the context that they don't need to explicitly say: for example, the location and time likely come from the device's GPS and clock respectively (though they may need to confirm these are correct with respect to what they are reporting). This could be considered a one-way transmission of information, but if we see it as a conversational interaction with an information-processing agent (a service in the middle of Figure~\ref{fig:d2d3} then the agent may query them for clarifications, corrections, additions, etc). All of which would be potentially valuable in turning the report into actionable information. For example, if these facts aren't provided, the agent could ask for details such as the vehicle registration, a description of the driver, whether the vehicle is stationary or moving (and, in the latter case, its heading), etc.

\subsection{Information Fusion}

Continuing the example, when the conversational interaction ends and the report is submitted, the analytic services may be able to fuse the newly provided information with data from other sources. For example, a particular vehicle registration from the report may result in a database query, which returns other recent (or not-so-recent) sightings of the same vehicle, individuals who are potentially associated with the vehicle, their known associates, etc. These additional items of information may be in a wide variety of forms, including sensor-collected data (e.g., still or video imagery), other human-provided reports and open source data (e.g., from social media). Fusion may involve a range of analytic capabilities, including image processing, pattern-matching, natural-language processing and text mining, and must take into account varying levels of quality-of-information (QoI). Again, this process may unfold as a conversation between computational agents, exchanging queries and responses. This conversation can occur concurrently with the first one: as soon as the eyewitness reports the vehicle registration, the query may indicate the vehicle is associated with a particular {\color{black} suspect individual (e.g. potential criminal, smuggler, or poacher)}, and a recent image of that {\color{black} suspect} individual may be sent to the eyewitness, asking if that individual is the driver or a passenger of the vehicle.  

\subsection{Sensor Tasking}

In the discussion so far, the emphasis has been on human provided and already-collected data. In many cases, as a conversational interaction unfolds, there will be a need to collect new information, for example, to corroborate or expand existing information. In our example, the sighting of a vehicle associated with a {\color{black} suspect} individual may trigger a need to track the vehicle (if it is moving) and/or gain a corroborating image of the driver. This may involve tasking or deploying sensor systems (a drone or roadside camera system, say), which will often require the intervention of a human decision-maker (on the right of Figure~\ref{fig:d2d3}). Even in cases where the sensors can act autonomously, there may still be a point at which a human in authority needs to be notified of what's happening. In our example, a {\color{black} security analyst} may be alerted (e.g., via a push notification) that a {\color{black} suspect} has potentially been sighted, and asked to confirm (or select) a sensing asset to gather more information. This interaction may be relatively complex, involving decisions on the appropriateness of various available assets (not least in terms of their QoI), the time and operational impacts of (re)tasking them, and the management and security policies governing their access and use. 

\vspace{0.25cm}

\noindent
Through these linked examples and use cases, a rich set of interactions can be seen between ``smart'' data sources, information-processing software agents and decision-makers. The sequence of examples were data-driven, beginning with an eyewitness report (Figure~\ref{fig:d2d1}). Stepping further back, however, the examples assume earlier stages of mission planning, patrol and sensor deployment, identification of {\color{black} suspect} individuals and so forth. These earlier activities began with a decision-maker's (i.e., mission commander's) requirements (Figure~\ref{fig:d2d2}). The conversational view (Figure~\ref{fig:d2d3}) is flexible enough to cover all of these bidirectional interactions. 

The remainder of this paper is organised as follows: Section~\ref{sec:appro} examines an approach to supporting these kinds of rich interactions among human and machine agents, building on previous work~\cite{Preece:2014}. Section~\ref{sec:expts} reports a pilot experiment using a conversational agent prototype with human users to perform simple crowdsensing tasks. Section~\ref{sec:related} discusses some related work and Section~\ref{sec:conc} concludes the paper.

\section{Human-Machine Conversations}
\label{sec:appro}

The interactions shown in Figure~\ref{fig:d2d3} comprise both human-machine and machine-machine exchanges. A {\it human-machine conversation} in the context of this paper is a sequence of messages exchanged between two or more agents where at least one party in each exchange is a machine. Choice of an appropriate form for the messages is a challenging problem: humans prefer ``natural'' forms such as natural language (NL) and images, but these forms are difficult for machines to process, leading to well-known problems of ambiguity and miscommunication. A compromise lies in the choice of a {\it controlled natural language} (CNL) --- a subset of a natural language with restricted syntax and vocabulary --- designed to be easily machine processable (with low complexity and no ambiguity) while also being human-readable and writable~\cite{CLCE2004}. Different CNLs have strong trade-offs in expressiveness, precision, simplicity, and naturalness~\cite{Kuhn:2014}. With training it is possible for human users to communicate directly in CNL. However, unrestricted natural language is more preferable to human users so the proposed approach is to use CNL for machine-machine interactions and allow humans the choice to use unrestricted NL or CNL. A machine communicating with a human should also have the choice of using CNL or a more natural form such as NL or images. 

Using a CNL as a common message form, even when there are no humans directly involved in the exchanges, avoids any need for translation to/from another technical representation while also making all communications human-friendly: for example, making copying-in of human users and any subsequent auditing much easier~\cite{fusion2012}. Many CNLs have been defined; here a form of controlled English known as ITA Controlled English (CE) is used, which seeks to balance trade-offs among expressiveness, precision, simplicity, and naturalness~\cite{Mott2010}. An example statement in CE syntax is shown below; this identifies an individual known to be a {\color{black} suspect}:

{\color{black}
\begin{verbatim}
  there is a person named p1 that is known as `John Smith' 
  and is a suspect.
\end{verbatim}}

\begin{figure}[b]
\centering
\includegraphics[width=0.36\textwidth]{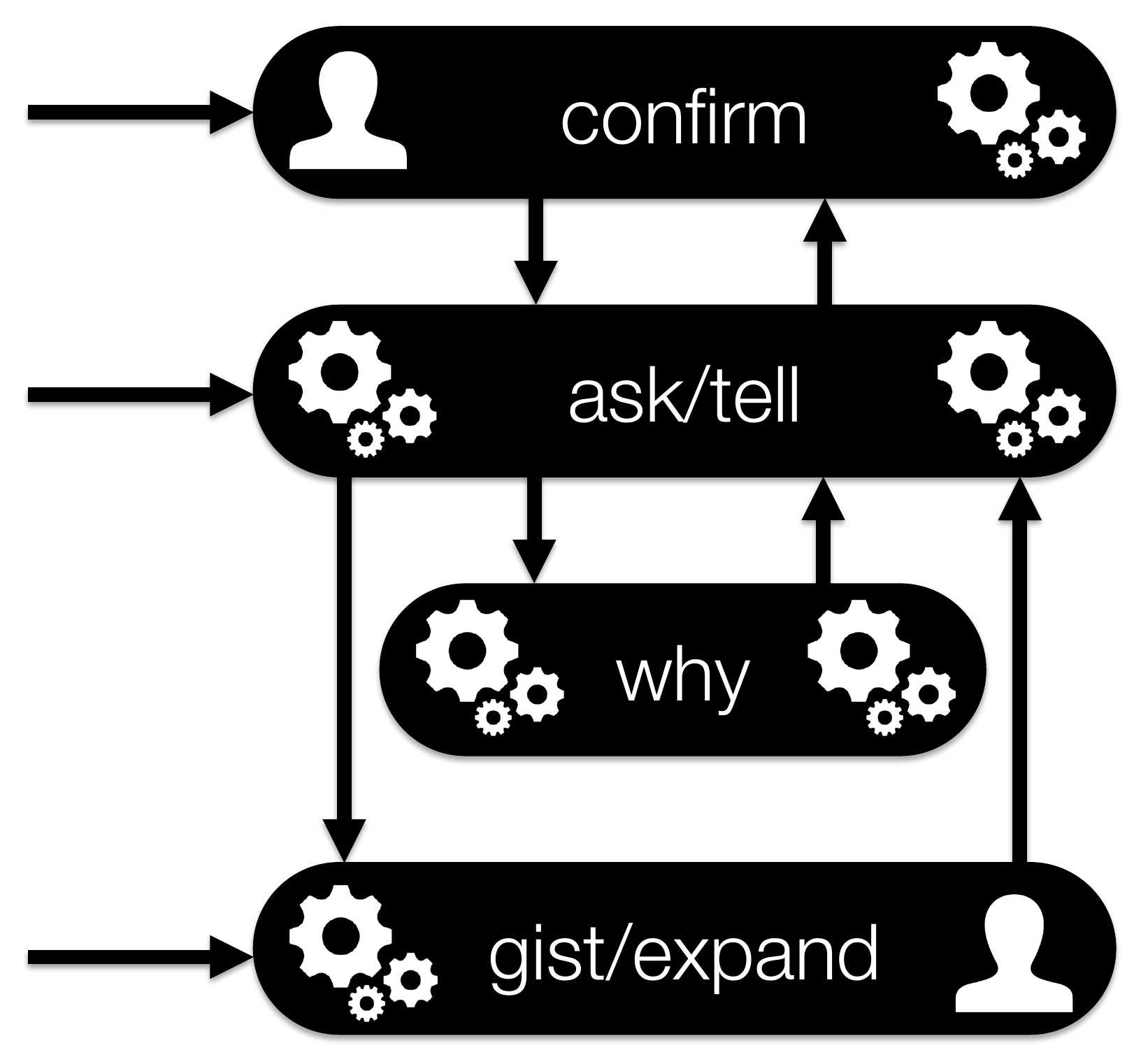}~~~~~~~~~~~~~~~\includegraphics[width=0.24\textwidth]{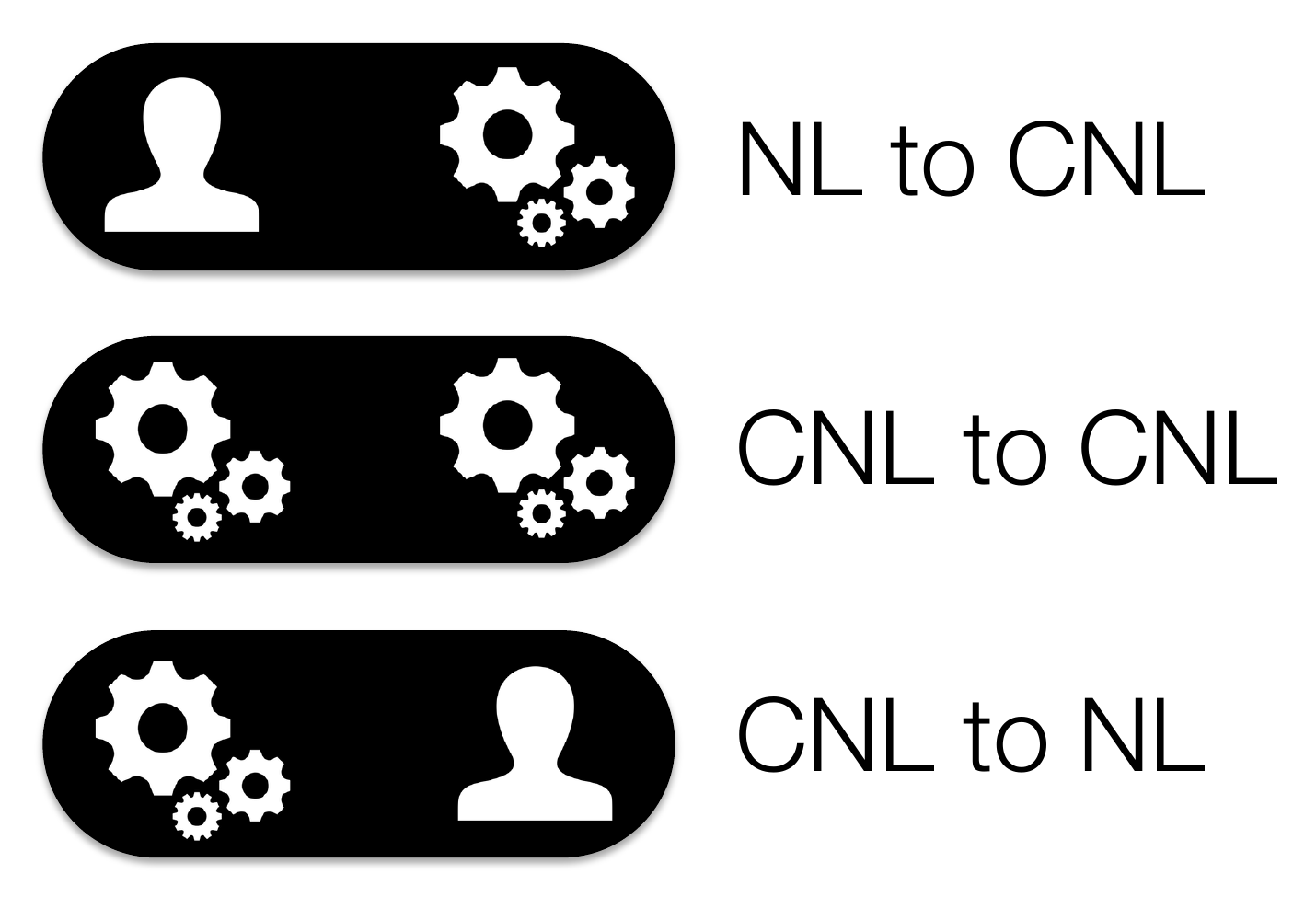}
\caption{Main types of conversational interactions and sequence}
\label{fig:sequence}
\end{figure}

Given the requirements to support human-machine and machine-machine dialogues, prior research in agent communication languages (ACLs) is relevant, where conversations were defined formally as sequences of communicative acts~\cite{fipa2002,Labrou:1997}. This work drew on earlier studies in philosophical linguistics: the idea of \textit{illocutionary acts} from speech act theory~\cite{Austin:1975} was adopted as a basis for ACL messages having explicit ``performatives'' that classify messages as, for example, assertives (factual statements), directives (such as requests or commands), or commissives (that commit the sender to some future action). The aim is to enable conversations that flow from natural language to CNL and back again through an exchange of messages. For this, a number of interaction types are required, detailed below. The interplay of these  is shown in Figure~\ref{fig:sequence}. (This is based on a more formal and complete treatment of the conversational protocol~\cite{Preece:2014}.) 
\begin{itemize}
\item A {\bf confirm} interaction that begins with a NL message (from a human user) and ends with a confirmed equivalent CNL form. Several steps may be involved in refining the user's intent to an acceptable CNL form. During the interaction --- which will usually be mediated by a conversational software agent --- there will often be negotiation of terminology (to reconcile the user's preferred terms with those used by the system) and the system may use natural forms of feedback including NL and images to achieve agreement on an acceptable CNL form of their message. Over time, as the user gains experience in interacting with the system, these exchanges should become shorter --- this is a hypothesis of the ongoing work reported in Section~\ref{sec:expts}. The confirmed CNL form of the user's intended message may be a query or a statement. The first use case in Section~\ref{sec:intro} --- spot report --- is an example of a confirm interaction.
\item An {\bf ask/tell} interaction that typically begins with a CNL query and ends with a CNL statement. Alternatively, these interactions may begin with a statement that leads to queries for additional information. Several agents may participate in an ask/tell interaction. The second and third use cases --- information fusion and sensor tasking --- are examples of ask/tell interactions. 
\item A {\bf gist/expand} interaction typically begins with a software agent attempting to communicate with a human by rendering CNL in a ``friendlier'' form such as a less restricted natural language format, or a combination of text and pictures. The term ``gist'' is used to capture the idea that these messages are intended to convey the gist of a more complex CNL message. In such cases, the recipient of a gist message may wish to request an expansion into full CNL, for example to clarify any ambiguity. Examples of gist/expand interactions appeared in the second and third use cases where humans (patrols, {\color{black} security analysts}) were sent notifications, for which the system is likely to choose a gist form.
\item An agent (human or machine) in receipt of CNL statements may wish to initiate a {\bf why} interaction to obtain the rationale for the information provided. The response will be a CNL ``because'' statement offering an explanation (for example, a summary of some reasoning or provenance for facts). In the examples, there are several places where this may be appropriate. For example, an analyst may request an explanation of why the system believes the vehicle is associated with the {\color{black} suspect}, or why a particular asset is being offered to track it (and perhaps why not some other asset).
\end{itemize}

\noindent
A conversational sequence can begin with any interaction except {\it why} and can then flow as indicated. The ``human'' and ``machine'' icons in the figure are intended to convey the kind of message forms involved in these exchanges, natural and CNL, rather than prescribing the kind of agents. As noted above, a key point of using a human-friendly CNL throughout the system is that it opens up all exchanges to human as well as machine participation.

\subsection{Use Cases Revisited}
\label{sec:coist}

Examples are given below of messages that might be exchanged in the context of each of the use cases from Section~\ref{sec:intro}.

\paragraph{Use case 1: Spot report}
A confirm interaction is initiated by a natural language message from the human patrol:
\begin{verbatim}
  Suspicious vehicle heading south: black saloon 
  with license plate DEF456
\end{verbatim}
The equivalent full CNL form for this in ITA Controlled English (CE) is:
\begin{verbatim}
  there is a vehicle named v48 that
   has DEF456 as registration and
   has the colour black as colour and
   has the vehicle body type saloon as body type and
   is a moving thing.
  there is a moving thing named v48 that
   has the direction south as direction of travel.
\end{verbatim}
The CE form uses a {\it model} of the world (also represented in CE) that defines concepts such as vehicle, moving thing, colour and direction, and relationships such as a vehicle having a registration, a colour and a body type, and a moving thing having a direction of travel. Note how certain terms have been negotiated in the conversation from the user's NL message to CNL: the user referred to a ``license plate'' which the system interpreted as ``registration'' in its model. Further details of how NL messages are interpreted into CE are given in Section~\ref{sec:expts}. In some cases, terms used by the user may not be interpretable in the model --- the word ``suspicious'' here has been ignored. In such cases, a more elaborate conversational interaction could be used to extend the model~\cite{Pizzocaro:2013} where appropriate. 

In practice, asking a user to confirm the full CNL shown above is challenging for several reasons. One such reason is because the CE form is relatively verbose so hard to read and also hard to display on devices with very limited screen sizes. We consider alternative ways to do this using the ``gist'' idea in the next subsection below.

\paragraph{Use case 2: Information fusion}

Following receipt of the user's confirmed CNL message, a fusion service infers the following CE:
{\color{black}\begin{verbatim}
  there is a suspect sighting named SS_v48 that
   has the vehicle v48 as target vehicle and
   has the person p1 as suspect candidate.
\end{verbatim}}
Note that the person (p1) referred to here is the subject of the example CE sentence given at the start of this section. In this way, a graph of interconnected facts is constructed. There are many possible recipients for this inferred information, not least the {\color{black} security authorities (e.g police, border agency, or environment protection officers)}. Parties who need to be informed of new {\color{black} suspect} sightings can ask to receive them and, in response, a fusion agent would tell this inferred fact to those parties (an ask/tell interaction). An agent in receipt of this fact may wish to obtain the rationale for the information, by engaging in a why interaction. The response, which includes supporting information, would be something such as:
{\color{black}\begin{verbatim}
  because there is a person named p1 
   that is known as `John Smith' and is a suspect and
   the person p1 has DEF456 as linked vehicle registration and
    there is a vehicle named v48 that has DEF456 as registration.
\end{verbatim}}

\paragraph{Use case 3: Sensor tasking}

The {\color{black} security authorities} may ask for the {\color{black} suspect} to be tracked using whatever means are available (this may be a standing request for all {\color{black} suspect} sightings). Previous work has addressed the use of CNL for representing sensing task requirements and matching these against a catalogue of available sensing assets~\cite{spie2012}. Using this approach, an agent can generate a tasking request as follows, and engage in an ask/tell interaction with an agent responsible for {\color{black} security} asset management  {\color{black} (e.g. deployment of unmanned aerial vehicles (UAVs)\footnote{a.k.a. ``drones''.} or ground camera systems)}:
\begin{verbatim}
  there is a task named TS_SS_v48 that
   requires the intelligence capability localize and
   is looking for the detectable thing car and
   is seeking instance the vehicle v48 and
   operates in the spatial area `North Road' and
   is ranked with the task priority High.
\end{verbatim}
Depending on the {\color{black} security} asset management protocols in place, a human may be notified of an assigned asset or asked to authorise an asset assignment. In either case, use of a gist/expand interaction would be appropriate. Here is an example notification in gist NL form:
{\color{black}\begin{verbatim}
  A MALE UAV with EO camera has been tasked to localize a black 
  saloon car (DEF456) with possible suspect John Smith in the 
  North Road area.
\end{verbatim}}
The gist/expand interaction could also be used to notify human patrols in the area (including the patrol that reported the original sighting in Case 1):
{\color{black}\begin{verbatim}
  Be on the lookout for a black saloon car (DEF456) with possible 
  suspect in the North Road area.
\end{verbatim}}
Here the D2D pipeline leads to action and, potentially, intervention/prevention. Note that all interactions are governed by information management policies. For example, the information relayed to patrols in this last step would depend on what they need --- and are permitted --- to know. In certain cases, some or all of this information could be withheld (e.g. they may be told to look out for the vehicle, but not informed of the link to a known {\color{black} suspect}). The conversational approach is well able to cope with such policy requirements, and CNL can be used to express the management policies as well~\cite{spie2013}.

\subsection{Prototype Conversational Agents}

\begin{figure}
\centering
\includegraphics[width=0.32\textwidth]{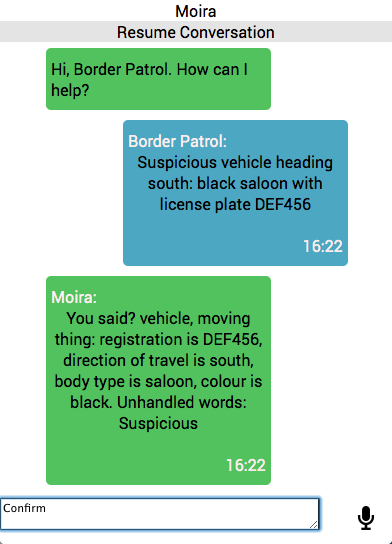}~\includegraphics[width=0.32\textwidth]{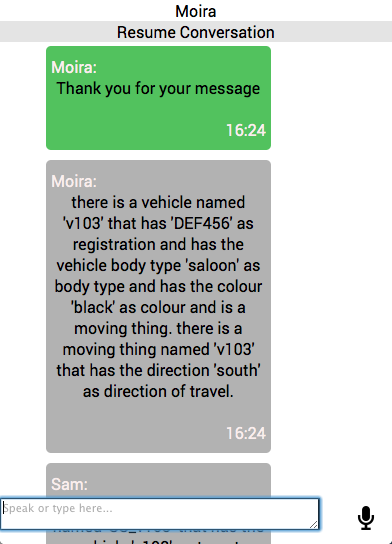}~\includegraphics[width=0.32\textwidth]{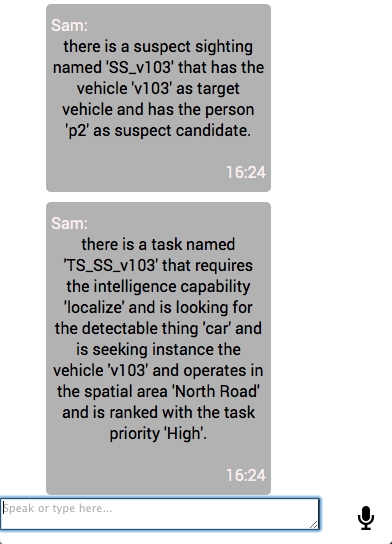}
\caption{A conversation with Moira and Sam agents using a prototype smartphone interface}
\label{fig:moira}
\end{figure}

To illustrate the conversational D2D concept, prototype conversational agents have been implemented and tested in limited experiments~\cite{Braines2014,Preece:2014}. Two distinct agent functionalities have been identified as useful and reusable:
\begin{itemize}
\item A conversational agent whose main purpose is to mediate interactions with human users (mainly confirm and gist/expand). This agent is called {\bf Moira} (Mobile Intelligence Reporting App).
\item A conversational agent whose main purpose is to apply knowledge of tasks and ISR assets to match tasks to available sensing assets. This agent is called {\bf Sam} (Sensor Assignment to Missions).
\end{itemize}

One interface to the Moira agent, deployed via a smartphone, is shown in Figure~\ref{fig:moira}. The sequence of screenshots depicted here reflects the three use cases described above. The smartphone user (whose name is {\color{black} ``Border Patrol''}) interacts with Moira by speech or typing. Their messages are shown in blue. Moira's messages are in green. In this case, the user is also permitted to see other conversations in which Moira is involved (shown in grey), so they see the exchange between Moira and Sam that initiates the new task request to track the vehicle. Note that the form of the confirmatory message shown in the second green bubble in the leftmost screenshot uses a gist form rather than full CE, for the reasons given above (brevity and low complexity). 

\begin{figure}
\centering
\includegraphics[width=0.54\textwidth]{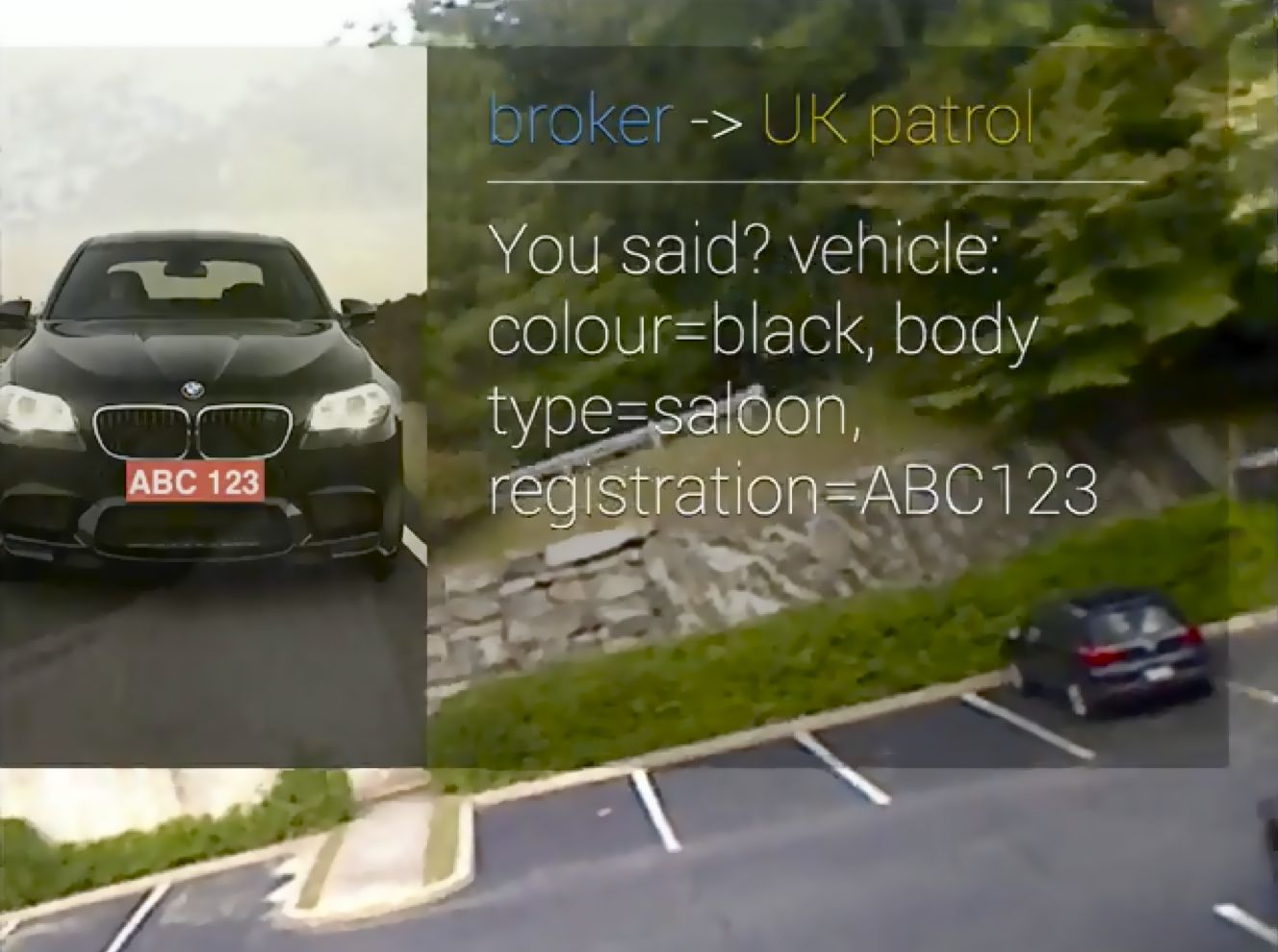}
\caption{Experimentation with a graphical form for confirmatory messages}
\label{fig:gist}
\end{figure}

A pilot implementation of Moira has also been created for an eyeline-mounted display such as Google Glass\footnote{http://glass.google.com}. Early experiments suggest a gist form of confirmatory message is even more appropriate here. An example of this is shown in Figure~\ref{fig:gist} where the user sees a combination of machine-generated images and compact text. In general, the style and format (e.g.,\ text and/or graphics) chosen by Moira for confirm and gist/expand interactions can be based on additional contextual factors such as the user, their role, the current operational tempo and the form factor of the device they are using.

\section{Crowdsensing Experiment}
\label{sec:expts}

A pilot experiment was conducted to test a prototype version of the Moira conversational agent. In keeping with Use Case 1 --- spot reports --- the experiment was designed as a crowdsensing exercise, in which subjects would view a series of scenes depicted in still images and describe them in natural language to the agent via a text-based interface. The agent would then provide feedback in CE on what it had understood from the user's input. To encourage users to submit more information, users were given a score for each submission, calculated in terms of the number of CE entities and relationships the agent was able to extract from the submission. For this experiment, subjects were 20 UK undergraduate students and the scenes depicted some common activities of emergency services (police, fire and ambulance) personnel. The main aims of this experiment were to (a) determine the degree to which the conversational agent could transform unrestricted NL descriptions into coherent CE, (b) test the robustness of the agent prototype with untrained users, (c) gain experience in providing a score-based feedback mechanism and (d) gather some baseline data on the usability of the prototype, to allow comparison with future planned studies. 

\subsection{Method}

Participants were issued with the following instructions:

{\sf You will be shown a sequence of images, each depicting a scene. For each one: 
Look at the picture on the screen. 
Describe the scene shown in the picture, using simple, concise English sentences 
e.g. the kind of sentences you might find in a book for children.
Submit each sentence one at a time, by pressing the 'Submit' button after completing each sentence.
Press 'Finish' when you have no more sentences to submit for the scene.

Don't worry about creating huge numbers of sentences
--- consider what you're trying to say and how the system handled previous sentences you submitted.
We're available to help if you need it
--- especially with explanations of bugs and misinterpretations.}

\begin{figure}
\centering
\includegraphics[width=0.8\textwidth]{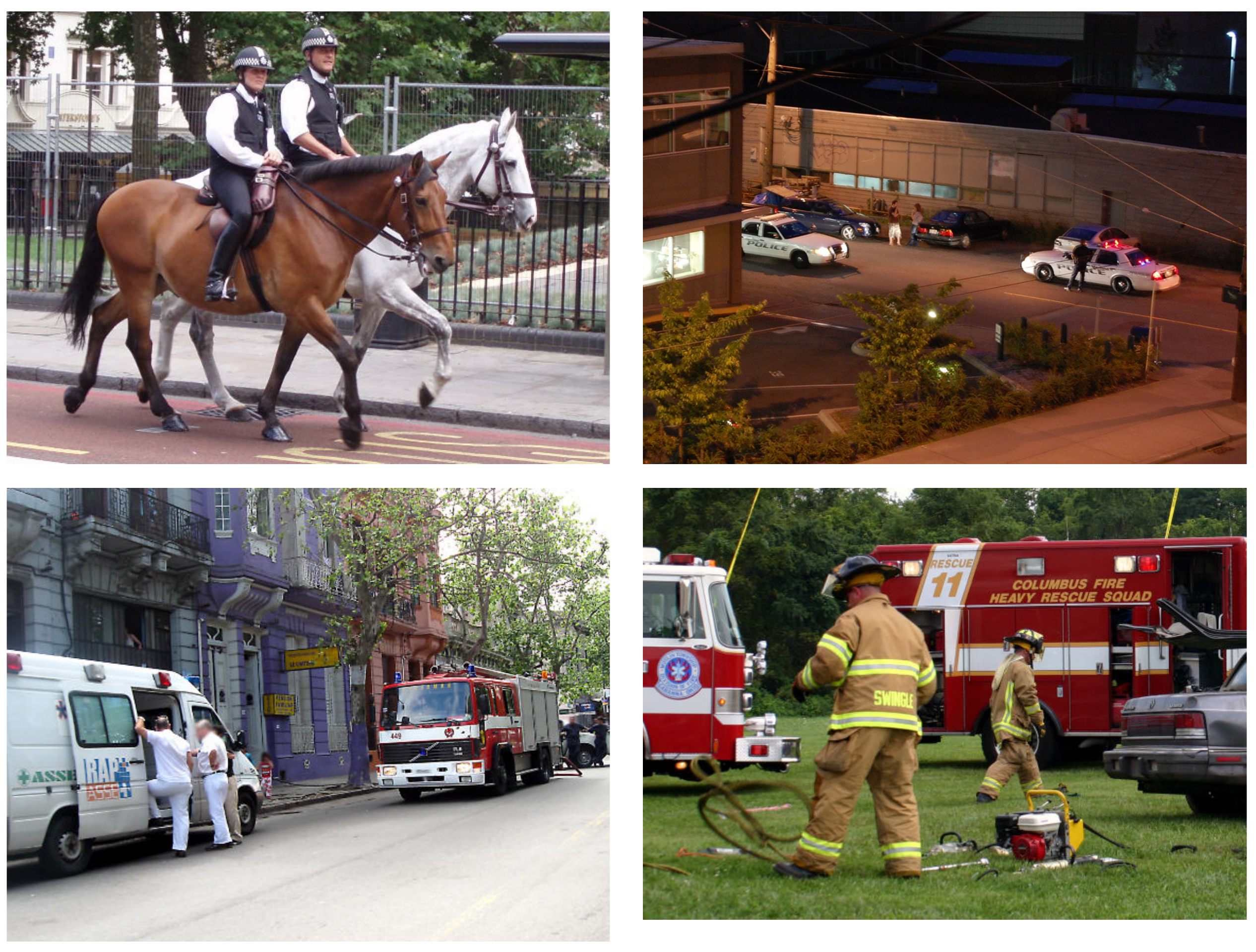}
\caption{Scenes used in the pilot experiment with the conversational agent prototype}
\label{fig:scenes}
\end{figure}

Four scenes --- Figure~\ref{fig:scenes} --- were shown to the group, each for 10 minutes. Feedback was given immediately after each submission in the form of the score and the CE recognised by the agent. The Simple Usability Scale\footnote{http://hell.meiert.org/core/pdf/sus.pdf} was used at the end of the exercise to obtain feedback.

\subsection{Implementing the Conversational Agent}
\label{sec:impl}

The conversational agent prototype used in the pilot experiment uses a {\it bag of words} approach rather than deep lexical parsing. This was motivated in part by the expectation that users will be trying to be helpful/straightforward with the system since it is in their interests in terms of getting better results if they do this. The incoming text is first split into phrases (multiple sentences), sentences, clauses (within a sentence, e.g., separated by commas or semicolons) and words. The agent then operates at a sentence level and scans the words left to right looking for matches to the CE model (details below). For each word (from left to right), the agent looks in the knowledge base to find any matches against the CE names of the concepts/instances/properties plus any synonyms defined (using CE). If a word is not matched (or even if it is), the agent then appends the next word and looks for matches again (to catch multi-word terms like ``police officer'' or ``is married to''). The agent progresses through all the words in the sentence and all matches for each word or phrase are aggregated against the words for later analysis. 
The way concepts and synonyms are modelled allows the user to decide the level of specificity they need. Some examples are shown below.

\noindent
{\it Defining concepts:} \\
\verb#  conceptualise a ~ vehicle ~ V.# \\
\verb#  conceptualise a ~ helicopter ~ H that is a vehicle.# 

\noindent
{\it Defining synonyms: } \\
\verb#  the entity concept `vehicle' # \\
\verb#   is expressed by the value `car' and # \\
\verb#   is expressed by the value `truck' and # \\
\verb#   is expressed by the value `sports car' and # \\
\verb#   is expressed by the value `bike'. # \\
\verb#  the relation concept `moving thing:direction of travel:direction' # \\
\verb#   is expressed by the value `driving' and # \\
\verb#   is expressed by the value `heading' and # \\
\verb#   is expressed by the value `going'. # 

\noindent
{\it Defining static common instances:} \\
\verb#  there is a colour named red. # \\
\verb#  there is a colour named black. # \\
\verb#  there is a colour named blue. # \\
\verb#  there is a direction named north. # \\
\verb#  there is a direction named south. # 

\noindent
Having aligned the words provided by the user to concepts, properties and instances in the model, the analysis of the meaning then takes place. Currently, the algorithm looks to see whether identified properties match the domain of the subject concept or instance.  If so they are matched against that subject. If not, then a separate CE sentence is generated. Where concepts are matched, new instances are generated with new ids (e.g., {\tt the helicopter h1\ldots})
Where named individuals are detected then that named individual is used (e.g., {\tt the person Fred\ldots}).
Detected properties use their range to determine what the subject type is (e.g.,``Fred is married to Jane'' is parsed to detect ``is married to'' as a property and since the range and domain are person then Fred and Jane are assumed to be instances of person). If they already exist as instances with that name then they will be used, otherwise a new instance with a temporary id such as {\tt p1} is generated, but the text used in the natural language sentence is used as a description (e.g., {\tt the person p1 is married to the person p2.} {\tt the person p1 has `Fred' as description.} {\tt the person p2 has `Jane' as description.}) An area for future work is to enable the agent to scan multiple sentences and incorporate anaphoric references and other patterns that span sentences.

\subsection{Results}

The experiment was run as an open session with a varying number of participants. Around 35 individuals participated. 137 scene descriptions were submitted in total. It was observed that the scoring mechanism (awarding 1 point for each CE concept, instance, or property recognised by the agent) had a positive effect on many of the participants, encouraging them to ``game the system'' and submit ever-more-elaborate descriptions. No attempt was made in this experiment to check the accuracy of submissions against ground truth. For example, a description of a ``blue horse'' or a ``green fire truck'' would gain the same points as a ``white horse'' or ``red fire truck''.

\begin{table}[h]
\centering
\begin{tabular}{|lrcrc|}
\hline
& Max & Min & Mean & Median \\
\hline
Phrases & 1 & 1 & ~1.00 & 1 \\
Sentences & 29 & 1 & ~1.92 & 1  \\
Clauses & 77 & 1 & ~3.26 & 1 \\
Words & 622 & 1 & 25.07 & 6 \\
Score & 19 & 0 & ~2.30 & 2 \\
\hline
\end{tabular}
\caption{Summary statistics for the 137 scene descriptions submitted by the 20 subjects}
\label{tab:stats}
\end{table}

\begin{figure}
\centering
\includegraphics[width=0.85\textwidth]{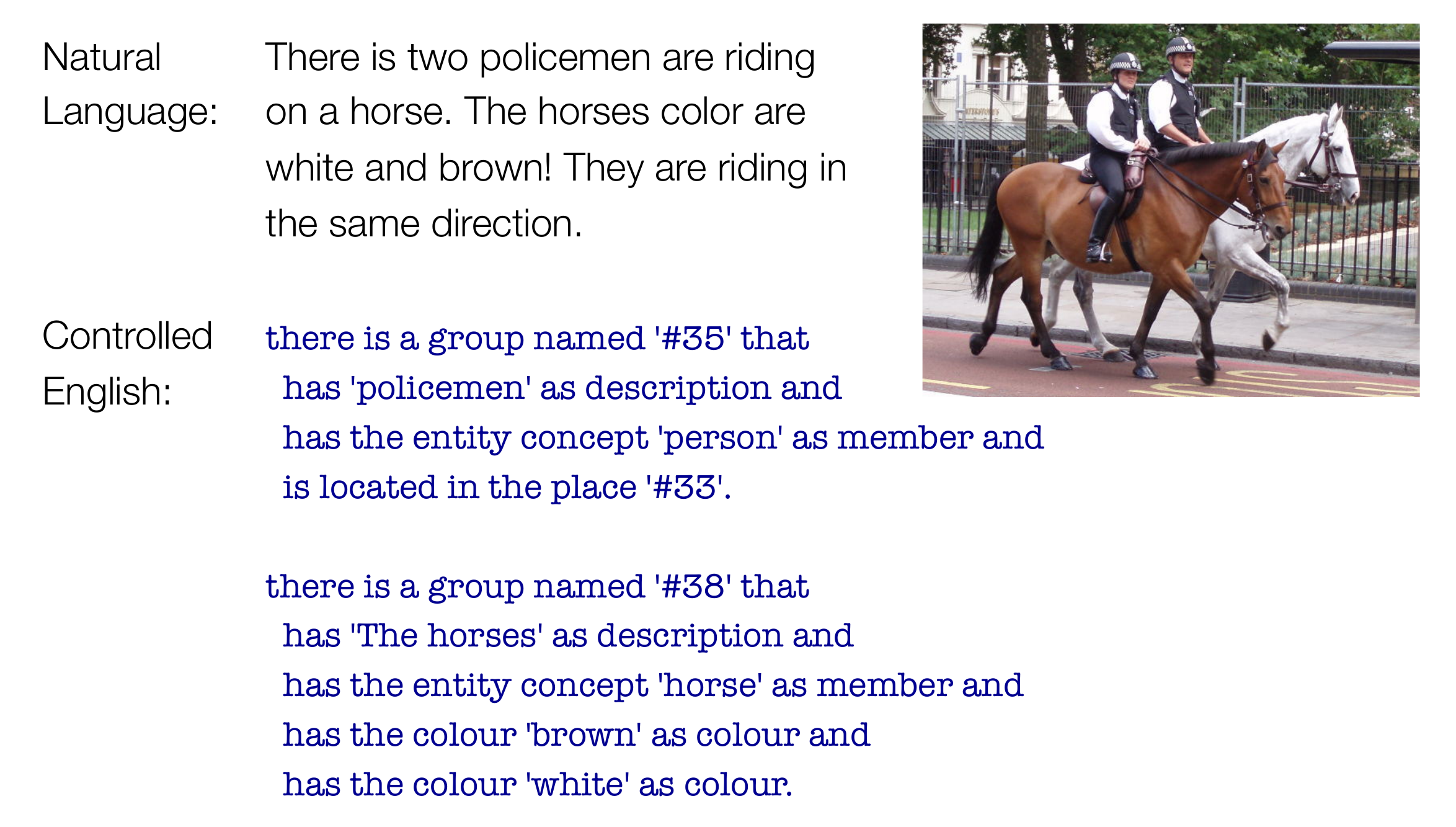}
\caption{Highest-scoring scene description submitted during the experiment (score$=$19)}
\label{fig:best}
\end{figure}

Table~\ref{tab:stats} summarises some statistics for the 137 scene descriptions. The data are skewed, so we give median and mean values. The longest description (622 words) was an example of a participant attempting to ``game'' the agent by submitting what amounted to a short piece of fiction describing the scene. The average score of 2.3 (median 2) was considered very acceptable: around two recognised model elements (concept, instance, or property) per description. The highest-scoring description is shown in Figure~\ref{fig:best}. The results from the Simple Usability Scale (out of 100) were: mean 64, standard deviation 13.13, maximum 95, and minimum 45. These were higher than expected given the early status of the prototype agent but do not mean much in isolation; as noted above, the main aim in gathering these was to provide a baseline for comparing future improved versions of the agent.

\section{Related Work}
\label{sec:related}

The high profile of intelligent language-understanding systems such as Siri and IBM's Watson\footnote{http://www.ibm.com/watson} have led to renewed interest in conversational interaction. Open problems include how to imbue machines with more natural conversational behaviours including turn-taking and user interruptions~\cite{Hastie:2013}, and how to operate effectively beyond static domains~\cite{Gasic:2013}, to reduce problems of brittleness common in these kinds of system. Mass-market intelligent agents such as Siri and Google Now remain essentially confined to simple ask-tell or tell interactions rather than flowing conversations.

The conversational approach is one type of human-computer collaboration (HCC), in which humans and intelligent systems work together with a common goal~\cite{Terveen:1995}. There is a growing body of HCC work in relation to collaborative intelligence analysis. {\color{black} Security analysts} are increasingly well versed in modern collaboration environments and social media, and systems are emerging that seek to combine the benefits of these approaches with existing software tools and processes for structuring and supporting tactical intelligence analysis. A recent example of this seeks to enable analysts to identify the decision-relevant data scattered among databases and the mental models of other personnel by employing familiar social media-style collaboration techniques~\cite{Wollocko:2013}. There is some evidence to indicate that, not only is it useful to collaborate within the same analyst team, but when collaboration is extended to the crowd and mediated by an intelligent software agent, the outcome of the intelligence analysis can be greatly improved~\cite{Brantingham:2013}. The authors propose a web-based application to collate imagery of a particular location from media sources and provide an operator with real-time situational awareness. Such approaches are promising, showing that a richly collaborative environment --- social, HCC, or both --- can be a blessing, if machines can help in sorting, filtering and managing large amounts of information. However, the same approaches can be a curse if the volume of information is simply increased.

\section{Conclusion \& Future Work}
\label{sec:conc}

This paper has set out a model for supporting flexible and agile D2D processes by means of a human-machine conversational approach. The model includes various kinds of interaction, including dialogues aimed at ameliorating ambiguity (confirm interactions), Q\&A dialogues (ask-tell), and explanatory dialogues (gist/expand, why). In the context of current concerns regarding transparency in big data approaches~\cite{Lazer:2014}, the latter kind of dialogues seem particularly promising.

Early results from experimentation with the conversational agent are encouraging. The prototype agent was able to extract useful CNL information from most of the NL inputs from untrained users. Going forward, the main aim is to improve the language understanding of the agent based on the CNL model and to support a wider range of conversational actions:
\begin{itemize}
\item From the user: questions, commands (to perform a task), multi-sentence narratives (building a story up) and model and lexical updates via NL (i.e., the addition of new knowledge of the kind shown in Section~\ref{sec:impl}).
\item For the machine: interjection (to seek knowledge) and requests for more details (e.g., if more information is needed to raise value or perform inference).
\end{itemize}

Another key aim is to extend the agent so that it is able to acquire input from more sources, for example audio/image/video input from the mobile device, metadata such as the device type/model, spatial and temporal data, and potentially even cues as to its user's emotional state. 

Three example forms of feedback from the agent to a human were considered in this paper: CNL, ``gist'' NL (see Figure~\ref{fig:moira}, leftmost screenshot) and a combined graphics/test gist formal (Figure~\ref{fig:gist}). Future experiments will focus on comparing these forms in terms of their effectiveness in confirm and gist/expand interactions. Further experiments will examine wider styles of conversation, general usability of the CE form of CNL, ability to quickly model or extend a model in a domain, multi-user conversations, and potentially also conversations with multiple different natural languages --- particularly important in coalition operations. 

The overall system of which the Moira and Sam agents form parts is highly modular and open to the integration of additional heterogeneous sources with relatively low effort. For example, the system could be extended to support crowdsourcing via social media by having the Moira agent operating behind a Twitter account so it can use Twitter to acquire information (either from public accounts or by asking), retweet etc. Social media also offers potential for building a highly distributed system with many different specialised agents.

Most of the discussion so far has focussed on the human interaction use case. There are also important problems to address in terms of conversationally-mediated information fusion. Handling more complex search queries potentially involving pattern analysis, and high-variety/volume/velocity data across space and time poses significant challenges, not least in semantic quality-of-information. Images/video are increasingly being tagged with meaningful pieces of information. Automated or semi-automated pattern analysis services offer increasing value in finding anomalies and potential threats. {\color{black} In relation to our specific examples, vehicles are often stolen for crimes}, so their travel pattern may be anomalous. {\color{black} If the criminals are wise they will steal from a different area to slow the law enforcement down to gain time to escape. So a security analyst} might wish to pose questions such as, ``Locations where vehicle with registration DEF456 has been spotted over the last month?'' or ``Number of times the vehicle with registration DEF456 has passed through the North Road checkpoint?'' Quantifying or qualifying the semantic QoI for responses to such questions is a hard open problem. 

Finally, the scope of this paper has been on D2D processes, but it is worth highlighting that the conversational approach does not stop at the point of decision: Section~\ref{sec:coist} showed an example where patrols were instructed to look out for the {\color{black} suspect}. Here we progressed from data-to-decision-to-{\it action}, affecting the state of the world. This is consistent with the view of conversations containing speech acts. The same thing happened in a machine-to-machine context where the Sam agent tasked a sensor. These examples show how feedback in a conversational system can improve future collection of data and refinement/enrichment of data already collected. This, in turn, raises interesting research questions as to how to optimise processes considering both machine and human sources.

\section*{Acknowledgements}
This research was sponsored by the US Army Research Laboratory and the UK Ministry of Defence and was accomplished under Agreement Number W911NF-06-3-0001. The views and conclusions contained in this document are those of the authors and should not be interpreted as representing the official policies, either expressed or implied, of the US Army Research Laboratory, the US Government, the UK Ministry of Defence or the UK Government. The US and UK Governments are authorized to reproduce and distribute reprints for Government purposes notwithstanding any copyright notation hereon.

\bibliographystyle{abbrv}
\bibliography{sensors}

\begin{thebibliography}{10}

\bibitem{Austin:1975}
J.~Austin and J.~Urmson.
\newblock {\em How to Do Things With Words}.
\newblock Harvard University Press, 1975.

\bibitem{Bakdash:2013}
J.~Z. Bakdash, D.~Pizzocaro, and A.~Preece.
\newblock Human factors in intelligence, surveillance, and reconnaissance: Gaps
  for soldiers and technology recommendations.
\newblock In {\em Proc MILCOM}, 2013.

\bibitem{Braines2014}
D.~Braines, G.~de~Mel, C.~Gwilliams, C.~Parizas, D.~Pizzocaro, and A.~Preece.
\newblock Agile sensor tasking for {CoIST} using natural language knowledge
  representation and reasoning.
\newblock In {\em Proc Ground/Air Multisensor Interoperability, Integration,
  and Networking for Persistent ISR V (SPIE Vol 9079)}. SPIE, 2014.

\bibitem{Brantingham:2013}
R.~Brantingham and A.~Hossain.
\newblock Crowded: a crowd-sourced perspective of events as they happen.
\newblock In {\em Proc Next-Generation Analyst (SPIE Vol 8758)}. SPIE, 2013.

\bibitem{Broome2012}
B.~Broome.
\newblock Data-to-decisions: a transdisciplinary approach to decision support
  efforts at {ARL}.
\newblock In {\em Proc Ground/Air Multisensor Interoperability, Integration,
  and Networking for Persistent ISR III (SPIE Vol 8389)}. SPIE, 2012.

\bibitem{Dumbill:2012}
E.~Dumbill, editor.
\newblock {\em Planning for Big Data}.
\newblock O'Reilly, 2012.

\bibitem{fipa2002}
{Foundation For Intelligent Physical Agents}.
\newblock {FIPA} communicative act library specification, 2002.

\bibitem{Fu:2006}
W.-T. Fu and W.~D. Gray.
\newblock Suboptimal tradeoffs in information seeking.
\newblock {\em Cognitive Psychology}, 52:195–2--42, 2006.

\bibitem{Gasic:2013}
M.~Ga\u{s}i\'ć, C.~Breslin, M.~Henderson, D.~Kim, M.~Szummer, B.~Thomson,
  P.~Tsiakoulis, and S.~Young.
\newblock {POMDP}-based dialogue manager adaptation to extended domains.
\newblock In {\em Proc SIGDIAL 2013}, pages 214--222, 2013.

\bibitem{Geyik:2013}
S.~Geyik, B.~Szymanski, and P.~Zerfos.
\newblock Robust dynamic service composition in sensor networks.
\newblock {\em IEEE Transactions on Services Computing}, 6(4):560--572, 2013.

\bibitem{Goldstein:2002}
D.~G. Goldstein and G.~Gigerenzer.
\newblock Models of ecological rationality: The recognition heuristic.
\newblock {\em Psychological Review}, 109:75–--90, 2002.

\bibitem{Hall:2007}
C.~C. Hall, L.~Ariss, and A.~Todorov.
\newblock The illusion of knowledge: When more information reduces accuracy and
  increases confidence.
\newblock {\em Organizational Behavior and Human Decision Processes},
  103:277–--290, 2007.

\bibitem{Hastie:2013}
H.~Hastie, M.-A. Aufaure, P.~Alexopoulos, H.~Cuay\'ahuitl, N.~Dethlefs,
  M.~Gasic, J.~Henderson, O.~Lemon, X.~Liu, P.~Mika, N.~B. Mustapha, V.~Rieser,
  B.~Thomson, P.~Tsiakoulis, Y.~Vanrompay, B.~Villazon-Terrazas, and S.~Young.
\newblock Demonstration of the {Parlance} system: a data-driven, incremental,
  spoken dialogue system for interactive search.
\newblock In {\em Proc SIGDIAL 2013}, pages 154--156, 2013.

\bibitem{Kuhn:2014}
T.~Kuhn.
\newblock A survey and classification of controlled natural languages.
\newblock {\em Computational Linguistics}, 40:121--170, 2014.

\bibitem{Labrou:1997}
Y.~Labrou and T.~Finin.
\newblock Semantics and conversations for an agent communication language.
\newblock In M.~N. Huhns and M.~P. Singh, editors, {\em Readings in agents},
  pages 235--242. Morgan Kaufman, 1998.

\bibitem{Laney:2001}
D.~Laney.
\newblock {3D} data management: Controlling data volume, velocity, and variety.
\newblock Technical report, META Group, 2001.

\bibitem{Lazer:2014}
D.~Lazer, R.~K.~G. King, and A.~Vespignani.
\newblock The parable of {Google Flu}: Traps in big data analysis.
\newblock {\em Science}, 343:1203--1205, 2014.

\bibitem{jdl:Llinas:2004}
J.~Llinas, C.~Bowman, G.~Rogova, A.~Steinberg, E.~Waltz, and F.~White.
\newblock Revisiting the {JDL} data fusion model {II}.
\newblock In {\em Proc Seventh International Conference on Information Fusion
  (FUSION 2004)}, pages 1218--1230, 2004.

\bibitem{Mott2010}
D.~Mott.
\newblock Summary of {ITA} {Controlled English}, 2010.

\bibitem{spie2013}
C.~Parizas, D.~Pizzocaro, A.~Preece, and P.~Zerfos.
\newblock Managing {ISR} sharing policies at the network edge using {Controlled
  English}.
\newblock In {\em Proc Ground/Air Multisensor Interoperability, Integration,
  and Networking for Persistent ISR IV (SPIE Vol 8742)}. SPIE, 2013.

\bibitem{Pizzocaro:2013}
D.~Pizzocaro, C.~Parizas, A.~Preece, D.~Braines, D.~Mott, and J.~Bakdash.
\newblock {CE-SAM}: A conversational interface for isr mission support.
\newblock In {\em Proc Next-Generation Analyst (SPIE Vol 8758)}. SPIE, 2013.

\bibitem{Preece:2014}
A.~Preece, D.~Braines, D.~Pizzocaro, and C.~Parizas.
\newblock Human-machine conversations to support multi-agency missions.
\newblock {\em ACM SIGMOBILE Mobile Computing and Communications Review},
  18(1):75--84, 2014.

\bibitem{Preece:2013}
A.~Preece, T.~Norman, G.~de~Mel, D.~Pizzocaro, M.~Sensoy, and T.~Pham.
\newblock Agilely assigning sensing assets to mission tasks in a coalition
  context.
\newblock {\em IEEE Intelligent Systems}, {Jan/Feb}:57--63, 2013.

\bibitem{spie2012}
A.~Preece, D.~Pizzocaro, D.~Braines, and D.~Mott.
\newblock Tasking and sharing sensing assets using controlled natural language.
\newblock In {\em Proc Ground/Air Multisensor Interoperability, Integration,
  and Networking for Persistent ISR III (SPIE Vol 8389)}. SPIE, 2012.

\bibitem{fusion2012}
A.~Preece, D.~Pizzocaro, D.~Braines, D.~Mott, G.~de~Mel, and T.~Pham.
\newblock Integrating hard and soft information sources for {D2D} using
  controlled natural language.
\newblock In {\em Proc 15th International Conference on Information Fusion},
  2012.

\bibitem{CLCE2004}
J.~Sowa.
\newblock {Common Logic Controlled English}, 2004.

\bibitem{Terveen:1995}
L.~Terveen.
\newblock Overview of human-computer collaboration.
\newblock {\em Knowledge-Based Systems}, 8(2):67–--81, 1995.

\bibitem{Wollocko:2013}
A.~Wollocko, M.~Farry, and R.~Stark.
\newblock Supporting tactical intelligence using collaborative environments and
  social networking.
\newblock In {\em Proc Next-Generation Analyst (SPIE Vol 8758)}. SPIE, 2013.

\end{thebibliography}

\end{document}